# Acquisition system for the CLIC Module.

S.Vilalte on behalf the LAPP CLIC group.[1]

1 – LAPP – CLIC group
Annecy – France

The status of R&D activities for CLIC module acquisition are discussed [1]. LAPP is involved in the design of the local CLIC module acquisition crate, described in the document STUDY OF THE CLIC MODULE FRONT-END ACQUISITION AND EVALUATION ELECTRONICS [2]. This acquisition system is a project based on a local crate, assigned to the CLIC module, including several mother boards. These motherboards are foreseen to hold mezzanines dedicated to the different subsystems. This system has to work in radiation environment. LAPP is involved in the development of Drive Beam stripline position monitors read-out, described in the document DRIVE BEAM STRIPLINE BPM ELECTRONICS AND ACQUISITION [3]. LAPP also develops a generic acquisition mezzanine that allows to perform all-around acquisition and components tests for drive beam stripline BPM read-out.

## 1    General specifications of a local acquisition system.

In CLIC, a large number of modules have to be acquired between the rare accesses from the surface (~900m). With a large number of channels, the CLIC module needs a local acquisition the closest as possible to the module. This choice is mandatory for technical and financial reasons (cabling, signal attenuation…). The module is a construction "brick" and its number of signals can fit a standard crate acquisition. So, the acquisition crate should be a part of the module, one crate for each module. Because of specific signals processing, costs and power consumption, the acquisition electronics should be « custom ». In another hand, it must support several module sub-systems, developed by several teams and which can evolve: connectivity & protocol have to be standardized.

The CDR document - *Accelerator control part* - describes the signals and specifications of the module [4]. The number of signals to acquire should be between 100 and 200 for the 7 sub-systems in the module. As each sub-system has different specifications and can evolve, or just can be present or not in the module, the more versatile solution is to implement a simple generic digital mother board able to receive mezzanines dedicated to sub-systems. Since 2008 LAPP works on the CLIC module acquisition in collaboration with CERN AB-BI-PI.

Designing at first a complete architecture is early because a lot of technical solutions have to be validated. A lot of inputs from sub-systems are also missing. Important choices on form factors or systems have still to be discussed with all the actors in the collaboration. The development began fall 2009 and several specifications were not fixed. So we proposed an architecture to the collaboration and we developed an evaluation prototype in parallel in order to validate technical choices.

## 2   Proposal of a future architecture of a local acquisition crate.

The most versatile solution should be to implement an industrial crate with a backplane able to receive:

- 1 service board: performs autonomous 12VDC power supplies from 230VAC line, performs the network access and its distribution on a back-plane.
- Several standard instrumentation motherboards: the architecture should be simplest as possible and should be based on FPGA and mezzanines. Different mezzanines should be developed for the different sub-systems with a standard interface (FPGA high speed connectors). Specific boards could also be developed according to the back-plane.

### 2.1   Network.

The network has to guarantee the data transmission on long distances with rates that can be important in some configurations (>100Mbps). It has also to take in charge the descending controls and clocks/timings synchronized with the machine. In some cases, synchronization is needed to avoid jitter on digitalization. It must be designed to resist to harsh environment: radiations, EMC.

Every crate should be linked point-to-point to the collection system: it limits hardware hard to access because of environment, it avoids intermediate stage in transfer, it isolates the crate if a problem occurs. In a complex system, it allows a better reliability.

Because of long distances, data rates, EMC and radiation environment, the best choice is an optical link between crate and collection (surface, alcove…).

The choice of a data collection system is larger because of the protected area where it should stand. We think a good solution should be a PCIe board because of its possible high data rates and because it is compatible with devices off-the-shell. It should be also a standard in the future. A synchronous rad-hard fiber optic network is developed in CERN microelectronics department for data acquisition, timing, trigger and experiment control: GBT [5].

This network should be the future LHC front-end standard network.

Features:

- 4.8Gb/s optical link.
- up to 40 local chip-to-chip links, ser-des (data concentration).
- Radhard design (~100Mrad, 1.E15  n/cm$^2$).
- multiple synchronous clocks management.
- Final version will include SCA slow control adapter (ADCs, DACs, JTAG, I2C, SPI, alarm monitoring…) very useful for the crate monitoring.
- HDL code developed for FPGA interfacing in the GBT collaboration.

## 2.2    Crate architecture.

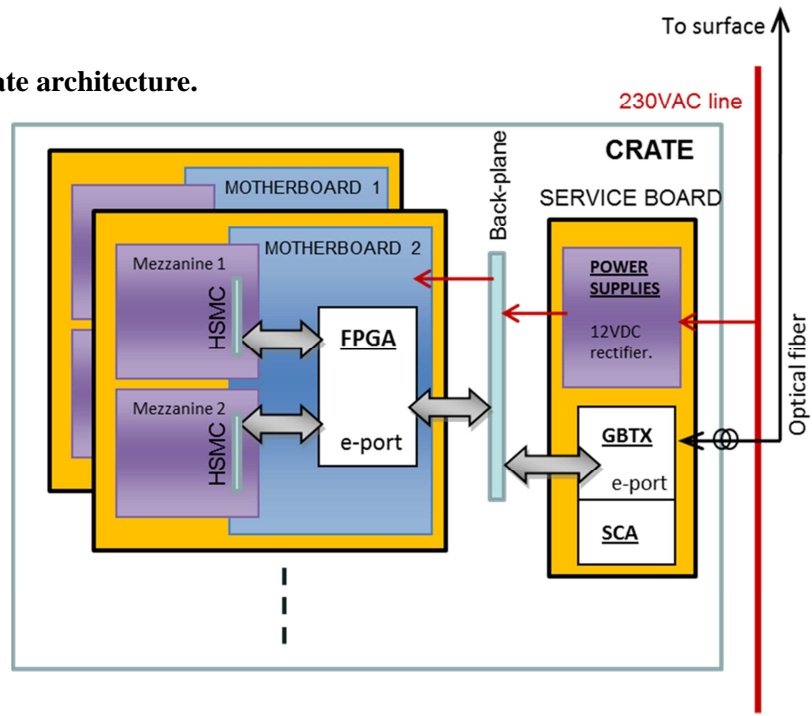

For better reliability, the crate should have the simplest architecture as possible.
Thanks to a custom back-plane, it should host the service board and several motherboards.
Front-end part of GBT distribution (e-links) should be cabled on the backplane, connecting motherboards to GBTX chip (ser-des and transceiver). Power supplies should be performed on the service board from AC line and distributed to motherboards by the back-plane too.
Motherboards should host two mezzanines. We think it is the best compromise between standard board sizes and possible front-panel connections for sub-systems signals.

On the standard instrumentation motherboard, a single FPGA (Altera ARRIA II) performs:
- interface with the GBT e-links via the backplane.
- interface with mezzanines dedicated to the different applications.
- possible implementation of a dedicated code for sub-systems applications: processing, feedback controls (attenuators, calibrations…).

We selected the very high speed Samtec HSMC connector for mezzanines-motherboards links.
- 8 transceivers links up to 8,5Gbps.
- 17 LVDS Tx, 17 LVDS Rx 1,2GBps.
- reserved links for clocks.

A motherboard with high speed mezzanines links will keep high performance in the future. It will allow upgrades for sub-systems mezzanines or future local applications developments with a standard connectivity (i.e. future high speed ADCs with serial high speed links transceivers).

### 2.3   Main issues.

#### 2.3.1   Radiations:

The main issue for a local acquisition is the radiation environment. The estimated total dose (TID) is about 1kGy per year so a total dose about 15kGy for a 15 years period + neutron fluence difficult to estimate. Digital components as FPGAs are the most sensitive. Specific Rad-hard FPGAs technologies are impossible to get because military or just not adapted for this application (number of cells, Rad-tol levels, speed…).
Technology and specific technics allow to limit radiation effects on industrial FPGAs:

- Small technologies are more resistant to TID and leakage currents anneal total    dose effects. TID effects decrease a lot for techno<90nm. SEL (single event latch-up) are limited by the size of the pnpn thyristor structure. Currently technologies are 40nm, next will be 28nm.
- Triple modular redundancy code techniques (TMR) fix the problems of non-destructive  SEU.
- A final hardcopy version, a kind of ASIC, should improve issues due to RAM susceptibility to neutron fluence. This solution is possible if no reconfigurable processing is needed in the FPGA.
- Shielding has to be studied and can reduce a lot the exposure.
- The use of industrial FPGAs for radiation environments is the current trend; a lot of information are available with the working group in CERN      (Jorgen  Christiansen fpga-radtol@cern.ch).

Analog components are rarely qualified for radiation use but are selected by experience and are known to be much more resistant. Concerning power supplies, CERN microelectronics group develops Rad-hard DC-DC converters (~3MGy, 5.E15n/cm²).
Finally, all parts should be qualified.

#### 2.3.2   Power consumption:

We should have low available power consumption for the whole crate, about 30-50W. Special care has to be taken in the design and the possibility to switch off the power supplies between two acquisitions has to be studied. This is not obvious for digital because of latencies and delays for PLLs locking.

# 3  Evaluation board, developments and prototypes.

## 3.1  Evaluation board.

Before crate prototyping, we developed an evaluation board to test architectures, GBT, mezzanine boards: it allows the emulation of a crate with two instrumentation boards and the emulation of four crates. This board will also be a future platform for developments and tests of mezzanines for applications.

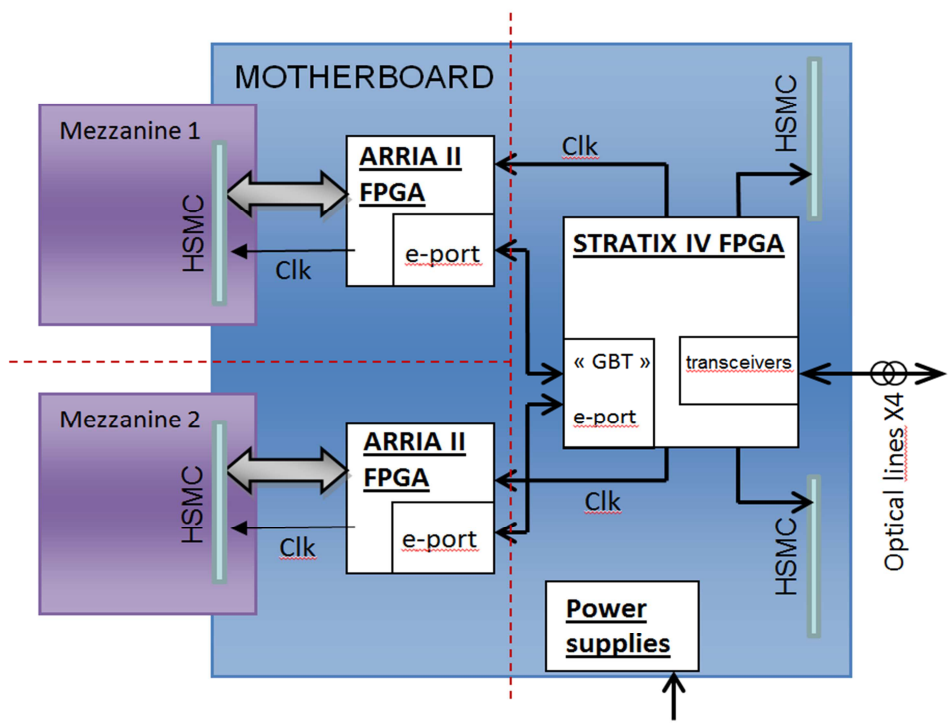

Architecture of the evaluation board with two mezzanines plugged.

The STRATIX IV FPGA emulates four GBT network links with four upstream optical links and local e-links with the two ARRIA II FPGAs. These two ARRIA II FPGAs emulate two instrumentation motherboards in the crate.

Two HSMC connectors will allow to test mezzanines including final GBT chip or power supplies. This architecture will allow to test all the functionalities, the mezzanine high speed links, the network and finally to make choices for the crate architecture.

This evaluation board has been produced beginning 2011:

240x240mm Hitachi FX-II
60µm class
18 layers
2000 components
900 differential pairs 1,2Gb/s
56 transceiver pairs 8,5Gbps
4 SFP 8,5Gb/s
3 FPGAs 1700 pins
4 USB links
4 mezzanines connectors

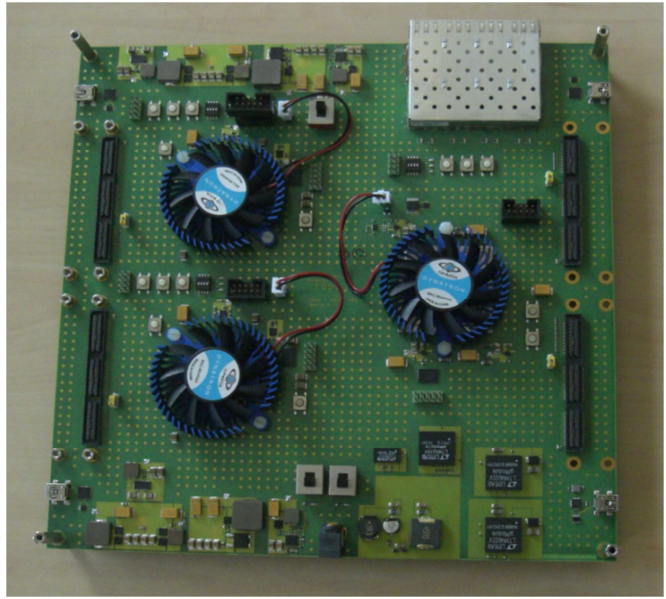

First tests gave good results for functionalities: HSMC connector tested with success beyond the specs (up to 10Gbps), all links between FPGAs, clock management and synchronization on board and between evaluation board and PCIe board via optical link.
GBT chip not yet tested.

## 3.2    Mezzanines developments.

In parallel of evaluation board, LAPP is involved in stripline BPM sub-system. We develop a specific mezzanine for stripline BPM shaping and acquisition in collaboration with CERN AB-BI-PI. A BPM acquisition needs 4 channels so quad ADCs are a good choice that allows to limit the dispersion (CMRR, synchronization…), that reduces the number of signals (clocks, controls, power supplie…) and that reduces the power consumption.
We developed at first a generic acquisition mezzanine:

- Allows to test components (amplifiers, ADC…) chosen for the stripline BPM acquisition.
- Can be useful for all around acquisitions and can be used to test first shaping circuits for BPMs or other sub-systems.

Drive Beam Stripline BPM specifications: (Steve Smith –SLAC- and CERN).
Resolution 2µm, accuracy 20µm, time resolution<10ns. Acquisition in baseband: 4-35MHz.
A shaping is needed with input filters.
Resolution needs high dynamic range acquisition: 75dB SNR, at least 100Msps.
Architecture needs multi-gain/attenuation to reach the resolution both in single pulse and train of pulses configurations. Finally, an accurate calibration is also needed.
Currently a design is studied by LAPP and CERN. Prototypes will be produced and tested beginning 2012. Beam tests are foreseen in 2012 in CTF3 DB & MB.

Generic acquisition board has been produced on spring 2011: 2 synchronous quad ADCs (Linear tech LTC2174)

ENOB=11,8; up to 125Msps
serial outputs. 2X4 channels
Possibility to implement gain (18dB)
Switchable attenuator (-26dB)
Clock jitter cleaner ~300fs
BW=50MHz
110x80mm
120 µm class
6 layers

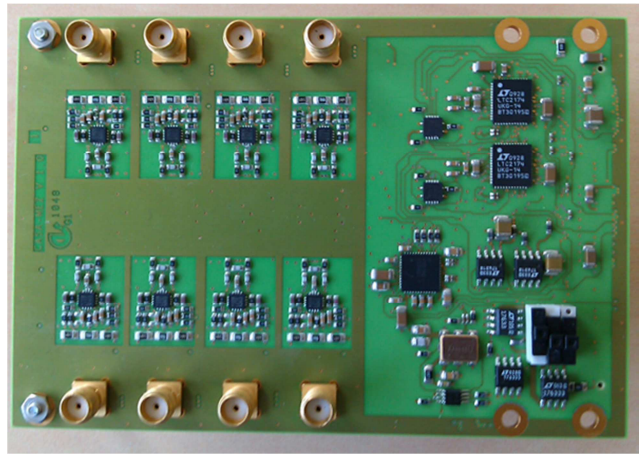

Generic acquisition mezzanine.

Dynamic tests have shown good results:
ADC: SNR=72dB, SINAD=71,2 so ENOB#11,5 @100Msps.
Amplifiers OK. Good HSMC connector transmission.
Good jitter cleaning and channels synchronization.

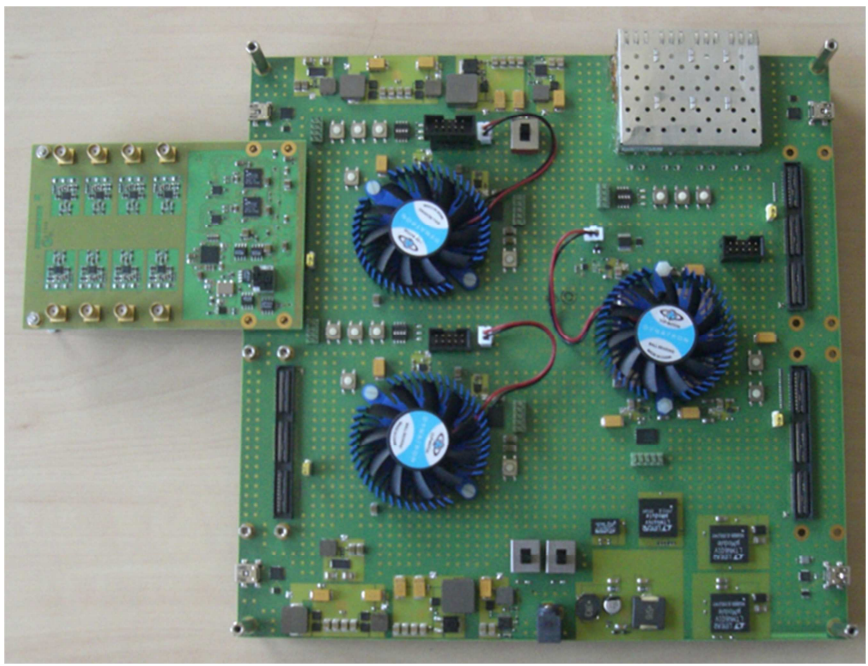

Generic acquisition mezzanine plugged on the evaluation board.

### 3.3     Data collection: X8 PCIe board.

In order to recover our prototype board data and in order to evaluate PCI express, we developed a PCIeX8 collection board. This board will not be necessarily the final chosen collection board, the aim is to evaluate its performances. PCIe will be the next bus standard and its advantage is to reach high speed transfers on standard platforms as PCs (direct memory access). This board allows to recover four 4,8Gb/s GBT optical links so four GBT links from four CLIC crates. The PCIe board will also be able to recover external machine clocks and timings and transmit them synchronously via GBT to the crates.
We produced a PCIe prototype on spring 2011:
- X8 Gen2 for high speed standard interface.
- STRATIX IV FPGA.
- 4 SFP+ optical links so recovers 4 crates.
- 2 MCX inputs for external trigger and clock: if external synchronization needed.
- 2 HSMC connectors allowing local mezzanine (i.e. generic acquisition)
- 1 USB link for tests and stand-alone use.

Results: all links tested up to 8,5Gbps. Acquisition tested with success.
Works well currently with USB, PCIe IP is still to be implemented (end 2011).

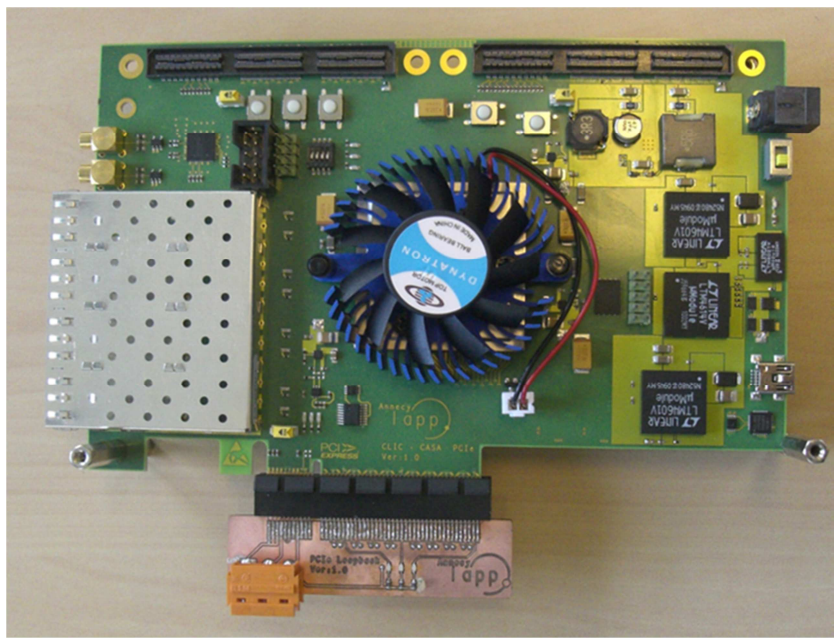

Possible global architecture of modules acquisition in a CLIC section:
Devices n are in a shielded area (surface or alcove). Green GBT links are optical links point-to-point between crates and PCIe boards. In a section there will be ~450 crates, so ~112 PCIe boards. It represents a reduced concentration system in term of occupation and power consumption.

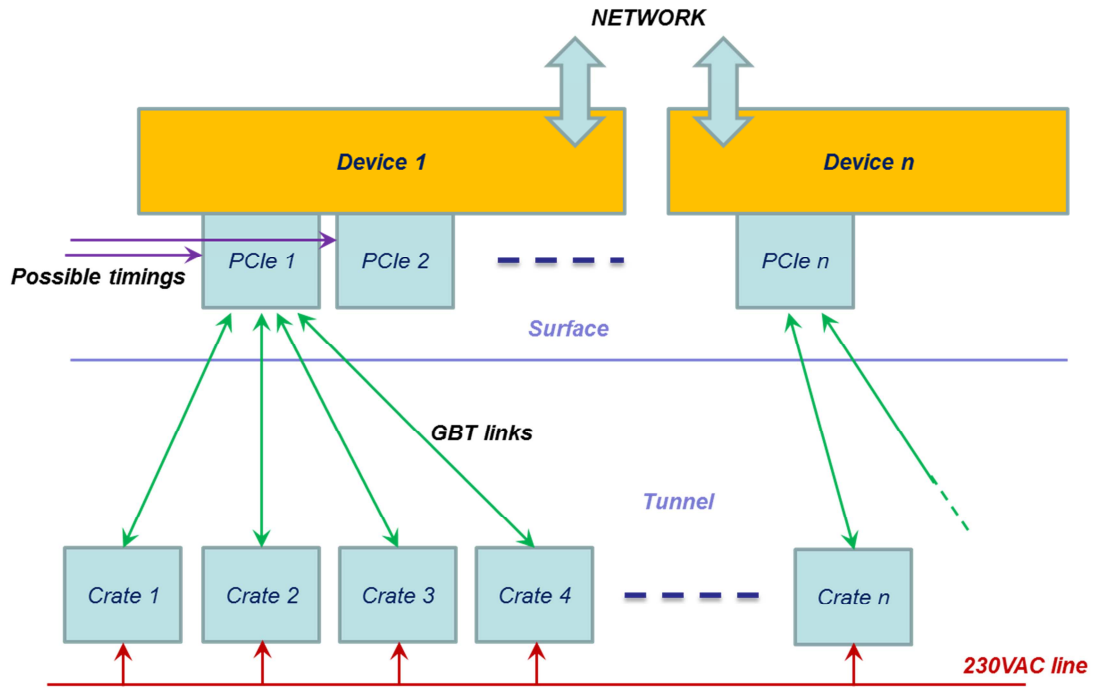

### 3.4    Perspectives.

Thanks to these prototypes, we are going to fix a prototype architecture for the crate in collaboration with CERN sections. In 2012 we plan to use our evaluation system to test stripline BPMs in lab and also in CLEX, the CTF3's experimental area.
We also plan to produce first prototypes for a full crate: mechanical crate, motherboard, service-board and stripline BPM mezzanine.